# Towards *de novo* RNA 3D structure prediction

Sandro Bottaro, Francesco Di Palma, and Giovanni Bussi

Scuola Internazionale Superiore di Studi Avanzati, International School for Advanced Studies, 265, Via Bonomea I-34136 Trieste, Italy

**Abstract**

RNA is a fundamental class of biomolecules that mediate a large variety of molecular processes within the cell. Computational algorithms can be of great help in the understanding of RNA structure-function relationship. One of the main challenges in this field is the development of structure-prediction algorithms, which aim at the prediction of the three-dimensional (3D) native fold from the sole knowledge of the sequence. In a recent paper, we have introduced a scoring function for RNA structure prediction. Here, we analyze in detail the performance of the method, we underline strengths and shortcomings, and we discuss the results with respect to state-of-the-art techniques. These observations provide a starting point for improving current methodologies, thus paving the way to the advances of more accurate approaches for RNA 3D structure prediction.

**Introduction**

RNA molecules carry out a large number of functions within cells, ranging from gene expression to gene regulation (i.e. riboregulation) and to catalysis. An increasing number of recent studies suggest that mutations in RNA-binding proteins and in non-coding sequences of the genome corresponding to RNA regulatory elements play an important role in many genetic diseases [1,2] such as in autoimmune diseases [3,4]. Additionally, changes in the level of non-coding RNAs have been observed in various cancers [5] and other pathological conditions such as cardiomyopathy [6] and neurodegeneration [7]. These diseases are typically associated with anomalous down- or up-riboregulation or with abnormal RNA molecules concentration [8]. Consequently, it is common to use specific RNA molecules as tumor markers [9] or for viral micro-RNA to be recognized as etiologic agents causing disease in humans [10]. Furthermore, RNA interference is considered as a new alternative in the therapeutic treatment of genetic and autoimmune diseases [11,12].

It is therefore of paramount importance to understand at a molecular level the function of RNA molecules. As for proteins, the function and mechanism of RNA molecules are intimately related to their three-dimensional structure – which is dictated by their sequence. A large suite of experimental approaches for determining the sequence and the secondary structure of RNA molecules exists. However, the atomic-detailed determination of the three-dimensional (3D) structure via X-ray crystallography or nuclear magnetic resonance (NMR) experiments of RNA is still very complex an expensive. This motivated the development of computational algorithms to model and predict RNA structure [13].

Many RNA 3D structure prediction algorithms typically rely on two ingredients: i) a sampling scheme, that generates putative RNA 3D structures (also called decoys), and ii) a scoring function, that, in the ideal case, makes it possible to identify among the decoys those with a native-like conformation.

In a recent publication, we have shown that relative nucleobase positions and orientations are sufficient to describe RNA structure and dynamics [14]. Additionally, we have introduced a framework that can be used to construct a knowledge-based scoring function for RNA structural prediction (ESCORE), which compares favorably to fully atomistic, state-of-the-art techniques.

In this research highlight we scrutinize the prediction capabilities of the ESCORE. We consider the cases in which our approach successfully and unsuccessfully predicts near-native RNA structures, and we analyze in details the 3D structure of the best predictions.

**Results**

The relative position and orientation of nucleobases in folded RNA molecules display very specific geometrical propensities, that are dominated by the presence of stacking and base-pairing interactions, as shown in Fig. 1.

To each point of the space around a nucleobase can be therefore assigned the probability of observing a neighboring base in that specific position and orientation. In other words, given a pair of bases, it is possible to quantify to which degree their relative position and orientation is compatible with the expected distribution observed in known RNA 3D structures. Since ribosomal RNA is the one for which the largest and most complex structures are available, we base our analysis on a high-resolution structure of the large ribosomal subunit (PDB code 1S72 [15]). For example, in this structure, observing two stacked bases, or two complementary bases forming a Watson-Crick base pair, is highly probable, while it is very unlikely to find two bases in very close contact (clashing). Recently, we have introduced a measure, called ESCORE, which makes it possible to quantify the accordance of any arbitrary RNA 3D conformation with the local, three-dimensional probability map obtained from the large ribosomal subunit shown in Fig. 1.

The ESCORE serves as a scoring function for RNA 3D structure prediction. More precisely, given a large number of conformations of an RNA molecule (decoy set), the ESCORE ideally ranks these conformations from the closest to the furthest from the native state, without prior knowledge of the native state itself.

We benchmark the ESCORE on 20 different decoy sets generated using the FARNA algorithm [16], which is a subset of the decoy sets we analyzed in our previous work [14]. This subset is particularly meaningful as a structure prediction exercise, since here the decoys were sampled without including any information about the actual experimental structure. For each decoy set, we assess i) if the known, native structure has the best score among all the decoys; and ii) the deviation between the native state and the best-scoring decoy. The latter test is more stringent, as in a real structure-prediction experiment the best-scoring decoy represents the putative native structure. The results, summarized in Table 1, show that for almost all decoy sets (18/20) the native state has a better (higher) ESCORE than any other structure in the decoy set (rank=0). Additionally, in these cases, the best-scoring decoy typically displays a similar secondary structure compared to the native state (Fig. 2), indicating that in a blind prediction test results would be satisfactory.

For two decoy sets, however, a non-zero number of decoys score better than the native structure. The most emblematic example is shown in Fig. 3. As in the previous case, the native and the best-scoring decoy both share the same secondary structure. However, we notice that the ESCORE relative to the stem region is higher in the decoy when compared to the native structure. By close inspection, we observe that this discrepancy is due to a significant difference in the base-base vertical distance between consecutive, stacked bases (3.66±0.53 Å for the decoy and 4.00±0.45 Å for native, considering the helix region only). More generally, it can be seen that the vertical distance between stacked bases in A-form helices of NMR models can deviate considerably from the typical distance observed in crystal structures. As an example, the vertical distance is 3.42±0.32 Å for the crystal structure of the large ribosomal subunit and 4.17Å ± 0.45 Å for the NMR structure of the nucleolin-binding RNA hairpin (PDB code 1QWA[17]).

This point is of particular importance when considering that ESCORE, as well as many other structure-prediction algorithms such as FARFAR [18] and RASP [19], are trained on high-resolution crystal structures. Therefore, it is not surprising that all these algorithms perform better on decoys relative to X-ray structures compared to NMR models, as also observed in the study of Bernauer *et al.* [20]

Finally, we inspect the results obtained on the challenging decoy set relative to the loop D/loop E arm of *E. coli* 5S rRNA, composed by a long stem with one terminal hairpin loop (PDB code 1A4D[21]). In this case, ESCORE correctly assigns to the native structure a good score (Fig. 4). However, the best scoring decoy is completely different from the native, featuring not one, but two hairpin loops. In Fig. 3 we also show that the secondary structure of the second-best scoring decoy is similar to that of the native state. Two items should therefore be addressed: i) why the two-best scoring decoys, that are completely different, have similar ESCORE; and ii) why is the best-scoring decoy so different from the native.

The answer to these questions can be found in the observation that the number of favorable base-base interactions in hairpin loops and in helices is very similar. Since the ESCORE can be considered as a weighted count of base-base interactions, good scores are assigned to both decoys, irrespectively of the presence of one or two hairpins. It is well known from secondary-structure studies that the presence of hairpin loops is energetically unfavorable with respect to helices [22]. This effect is not explicitly taken into account in the ESCORE, and is thus an important ingredient that could be used to improve the method. This explanation is also supported by the fact that the very same issue affects other scoring functions that heavily rely on local-contacts, such as FARFAR [18] and RASP [19].

**Conclusions**

The ESCORE can correctly discriminate and rank decoy structures among most of the cases examined in this Paper. When compared with state-of-the-art, all-atom methods such as FARFAR or RASP, ESCORE consistently performs equally well and in some cases better, thus demonstrating the validity of the approach.

We examined in detail two decoy sets for which ESCORE (as well as FARFAR and RASP) fails in identifying the closest-to-native decoy or in assigning the best score to the native structure, and we identified two major sources of errors. Firstly, we observed that it is more difficult to obtain good scores for RNA structures solved by NMR. This is likely due to significant differences in the helix geometry between X-ray crystal structures, which are used to parameterize ESCORE, and NMR models.

Furthermore, we have shown that the ESCORE does not penalize to a sufficient degree the presence of hairpins with respect to helices. This problem, affecting ESCORE as well as other state-of-the-art, atomistic methods, is an important shortcoming that should be addressed properly in the future in order to allow for accurate RNA 3D structure predictions.

**Acknowledgments**

This work was supported by the European Research Council under the European Union's Seventh Framework Programme (FP/2007-2013) [306662, S-RNA-S].

**Figure 1**. Three-dimensional distribution of nucleobases obtained from the crystal structure of the large ribosomal subunit. Different colors correspond to the different interaction types: Watson-Crick pairs in red/orange, non-canonical interactions in blue, stacked pairs in green.

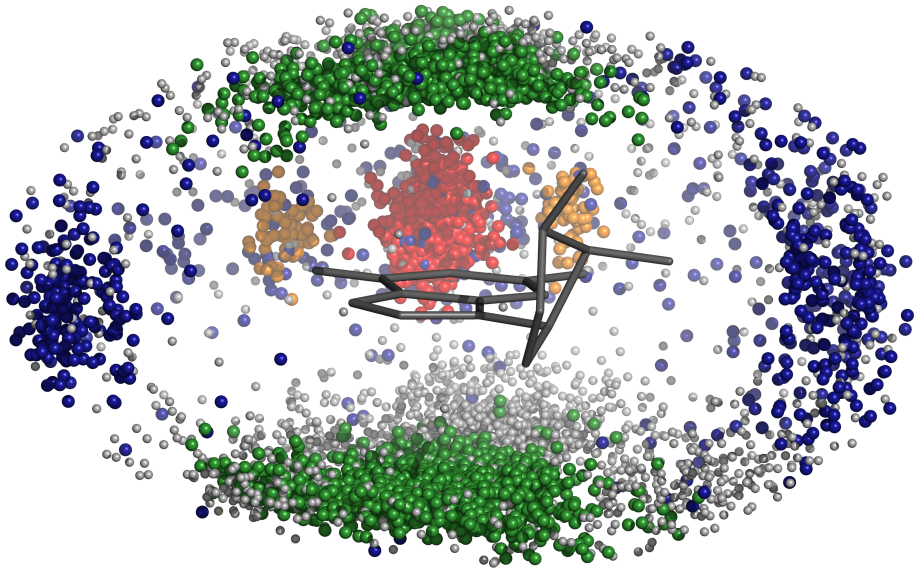

**Table 1**. Summary of ranking and decoy screening capabilities of ESCORE. For each decoy set, we report a) the number of decoys scoring better than native, over a total of 500 structures per decoy set; b) the RMSD from native for the best-scoring decoy; c) the distance of the closest-to-native structure within the decoy set. PDB codes relative to solution NMR structures are highlighted in gray.

| PDB code | Rank[a] | RMSD best (Å)[b] | Min RMSD (Å)[c] |
|---|---|---|---|
| 157D | 0 | 3.54 | 1.46 |
| 1A4D | 0 | 23.57 | 4.06 |
| 1CSL | 0 | 3.96 | 3.28 |
| 1DQF | 0 | 3.04 | 2.39 |
| 1ESY | 0 | 3.70 | 2.75 |
| 1I9X | 0 | 4.79 | 3.0 |
| 1J6S | 0 | 10.66 | 2.94 |
| 1KD5 | 2 | 4.22 | 2.52 |
| 1KKA | 151 | 4.15 | 3.53 |
| 1L2X | 0 | 13.62 | 3.81 |
| 1MHK | 0 | 9.03 | 4.68 |
| 1Q9A | 0 | 4.74 | 4.22 |
| 1QWA | 0 | 4.24 | 3.52 |
| 1XJR | 0 | 8.81 | 6.94 |
| 1ZIH | 0 | 1.84 | 1.63 |
| 255D | 0 | 1.90 | 1.72 |
| 283D | 0 | 3.12 | 1.73 |
| 28SP | 0 | 3.73 | 2.80 |
| 2A43 | 0 | 4.78 | 4.52 |
| 2F88 | 0 | 3.87 | 2.74 |

**Figure 2**. Native structure (left) and best scoring decoy structure within the 28SP decoy set. Each residue is colored according to the ESCORE. The sum of all per-residue contributions gives the total ESCORE value. Note that bases in the stem regions typically have high ESCORE values. Conversely, bases in the hairpin and internal loop regions have low ESCORE values. Root-mean-square deviation (RMSD[23]) from native of the best scoring decoy is also shown.

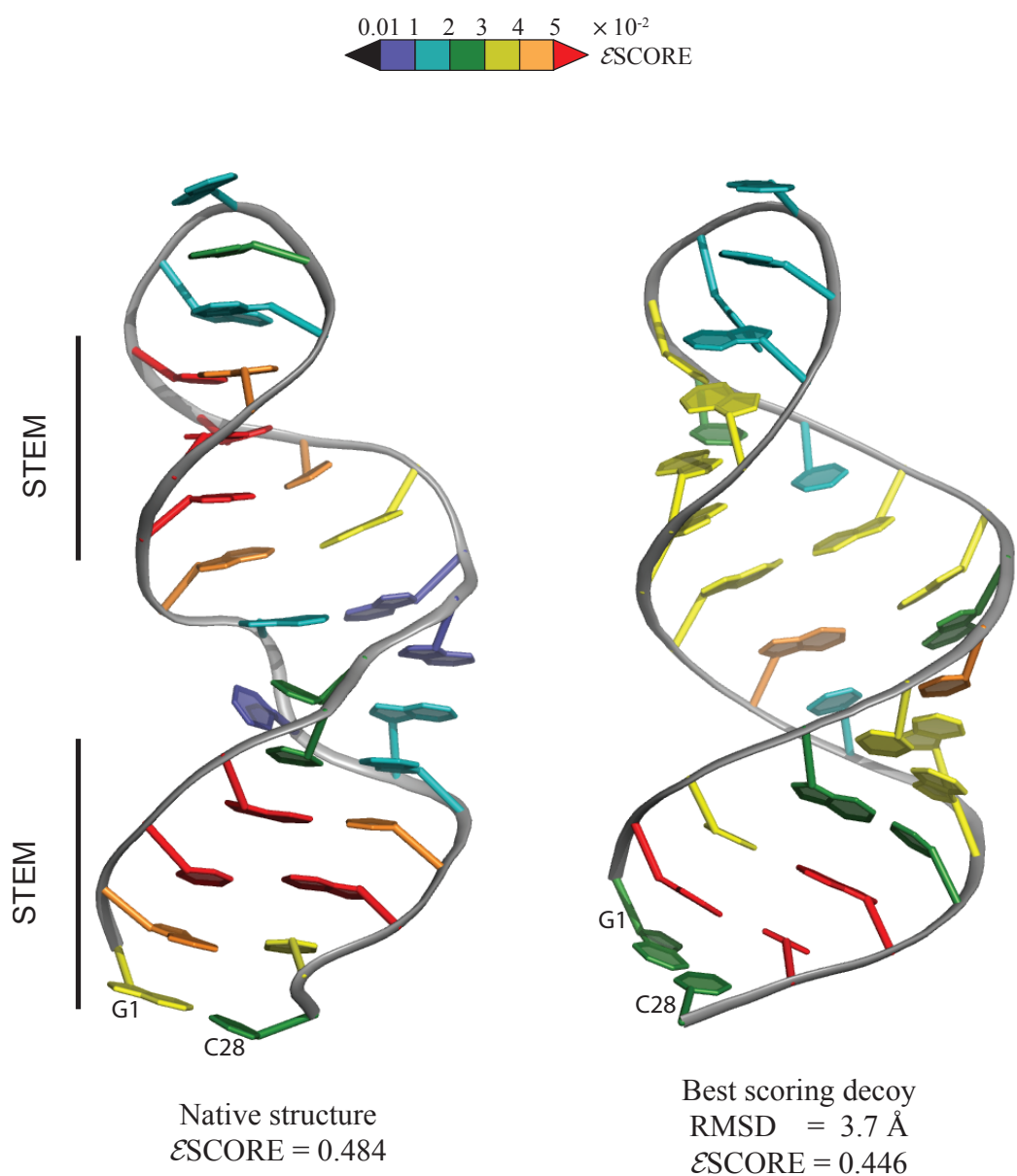

Native structure
$\mathcal{E}$SCORE = 0.484

Best scoring decoy
RMSD = 3.7 Å
$\mathcal{E}$SCORE = 0.446

**Figure 3**. Native structure (left) and best scoring decoy structure within the 1KKA decoy set. Each residue is colored according to the ESCORE. Note that the residues in the stem give the dominant contribution to the ESCORE. RMSD from native of the best scoring decoy is also shown.

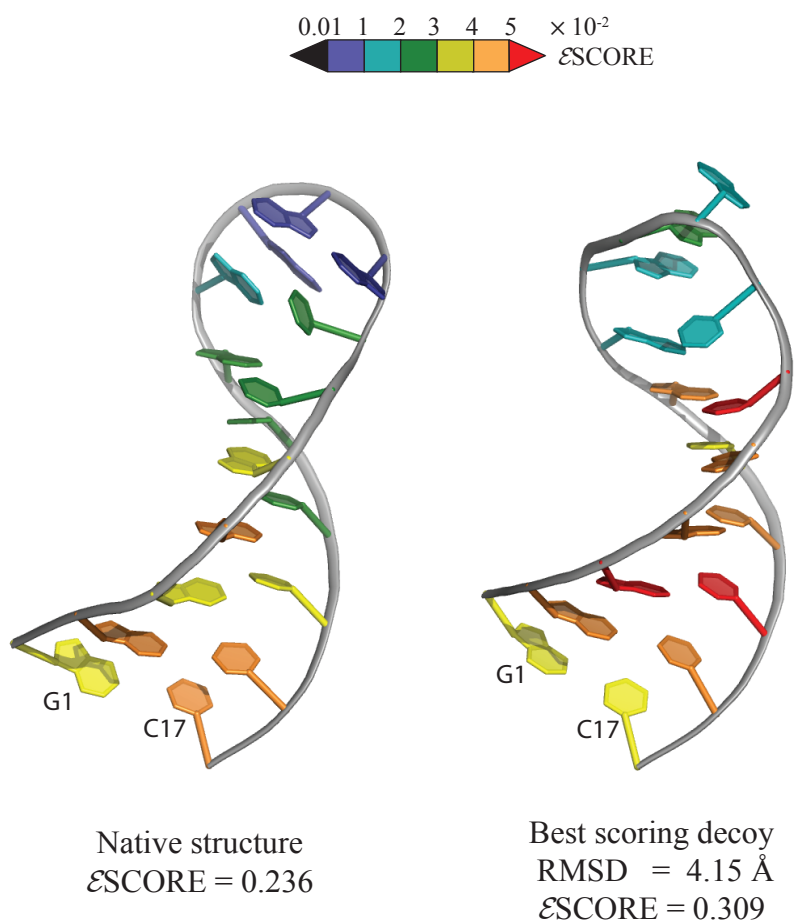

Native structure
$\mathcal{E}$SCORE = 0.236

Best scoring decoy
RMSD = 4.15 Å
$\mathcal{E}$SCORE = 0.309

**Figure 4**. Native structure (left) and best scoring decoys structure within the 1A4D decoy set. Each residue is colored according to the ESCORE. RMSD from native of the decoys are also shown.

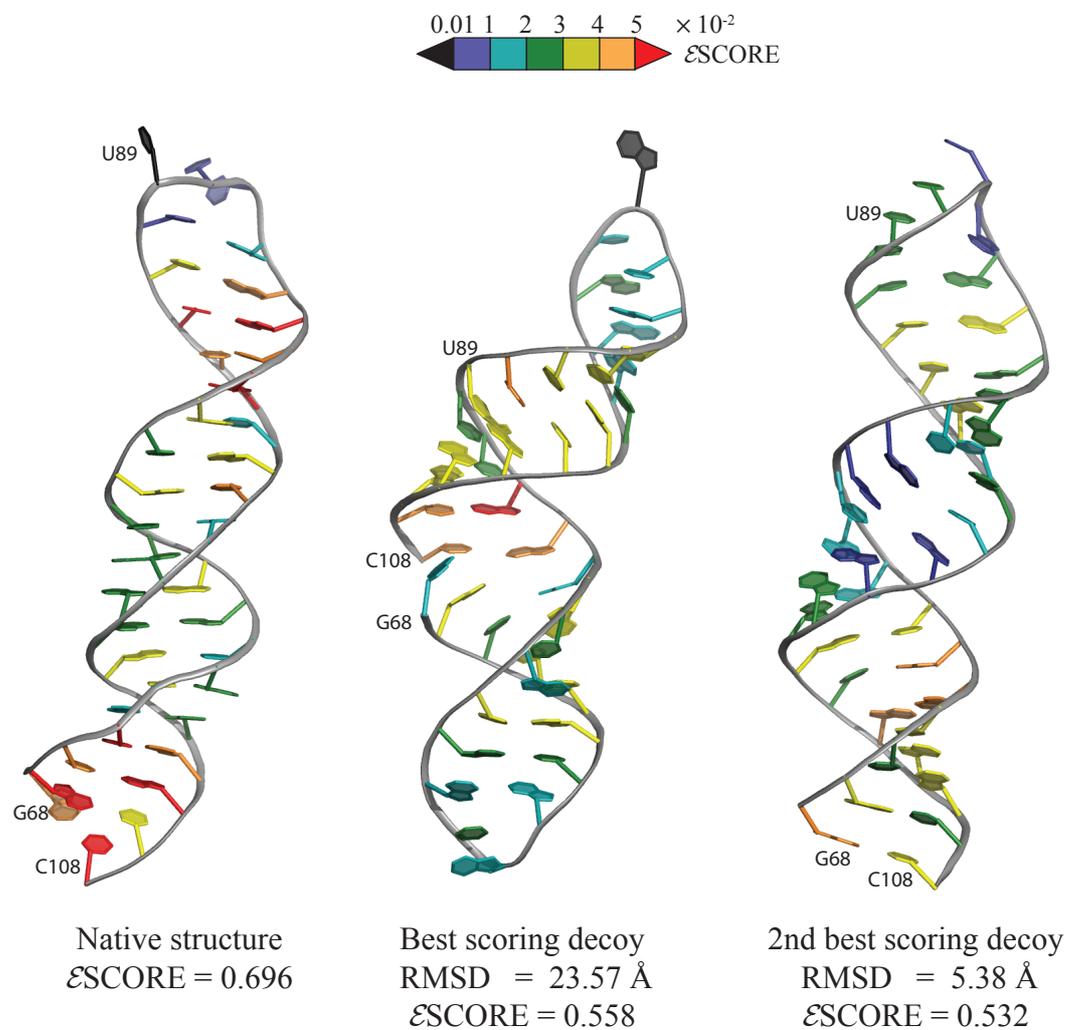

Native structure
𝓔SCORE = 0.696

Best scoring decoy
RMSD = 23.57 Å
𝓔SCORE = 0.558

2nd best scoring decoy
RMSD = 5.38 Å
𝓔SCORE = 0.532